\documentclass{PoS}

\usepackage{amsmath}
\usepackage{cite}

\newcommand{\gamtt}[4]{{\widetilde{\Gamma}}\!\left(\begin{smallmatrix}#1\\#2\end{smallmatrix};#3,#4\right)}

\title{Elliptic polylogarithms and two-loop Feynman integrals}

\ShortTitle{Elliptic Polylogarithms}

\author{Johannes Broedel,$^a$ \speaker{Claude Duhr},$^{bc}$
Falko Dulat,$^d$ Brenda Penante$^b$ and Lorenzo Tancredi$^b$\\
\llap{$^a$} Institut f\"{u}r Mathematik und Institut f\"{u}r Physik,
Humboldt-Universit\"{a}t zu Berlin,\\
IRIS Adlershof, Zum Grossen Windkanal 6, 12489 Berlin, Germany\\
\llap{$^b$} Theoretical Physics Department, CERN, Geneva, Switzerland\\
\llap{$^c$} Center for Cosmology, Particle Physics and Phenomenology (CP3),\\
Universit\'e Catholique de Louvain, 1348 Louvain-La-Neuve, Belgium\\
\llap{$^d$} SLAC National Accelerator Laboratory, Stanford University, Stanford, CA 94309, USA\\

        E-mail: \email{jbroedel@physik.hu-berlin.de},
\email{claude.duhr@cern.ch},
\email{dulatf@slac.stanford.edu},
\email{b.penante@cern.ch},
\email{lorenzo.tancredi@cern.ch}}

\abstract{We review certain classes of iterated integrals that appear in the computation of Feynman integrals that involve elliptic functions. These functions generalise the well-known class of multiple polylogarithms to elliptic curves and are closely related to the elliptic multiple polylogarithms (eMPLs) studied in the mathematics literature. When evaluated at certain special values of the arguments, eMPLs reduce to another class of special functions, defined as iterated integrals of Eisenstein series. As a novel application of our formalism, we illustrate how a class of special functions introduced by Remiddi and one of the authors can always naturally be expressed in terms of either eMPLs or iterated integrals of Eisenstein series for the congruence subgroup $\Gamma(6)$.}

\FullConference{Loops and Legs in Quantum Field Theory (LL2018)\\
		29 April 2018 - 04 May 2018\\
		St. Goar, Germany}

\begin{document}

\section{Introduction}

Progress in the analytic computation of Feynman integrals has always gone hand-in-hand with a deeper understanding of the class of mathematical functions and numbers that show up in perturbative Quantum Field Theory. A major breakthrough was achieved by Remiddi and Vermaseren almost 20 years ago when they realised that large classes of Feynman integrals that depend on a single ratio of scales can be expressed in terms of a new class of functions dubbed \emph{harmonic polylogarithms} (HPLs)~\cite{Remiddi:1999ew}. Since then it has become clear that in order to describe multi-loop and multi-scale Feynman integrals generalisations of HPLs need to be introduced~(see, e.g., ref.~\cite{Gehrmann:2000zt,Aglietti:2004tq,Bonciani:2010ms,Ablinger:2011te} for a non-exhaustive list). Many of these generalisations fall into the realm of a class of special functions going under the name of \emph{hyperlogarithms} or \emph{multiple polylogarithms} (MPLs) in the mathematics literature (cf., e.g. ref.~\cite{Goncharov_MPLs,Kummer,Lappo}). They are defined as,
\begin{equation}\label{eq:MPLs_def}
G(a_1,\ldots,a_n;x) = \int_0^x\frac{dt}{t-a_1}\,G(a_2,\ldots,a_n;x)\,,
\end{equation} 
where the recursion starts with $G(;x)=1$. If $a_n=0$ then the integral in eq.~(\ref{eq:MPLs_def}) diverges, and we define instead
\begin{equation}
G(\underbrace{0,\ldots,0}_{n\textrm{ times}};x) = \frac{1}{n!}\log^nx\,.
\end{equation}

MPLs are the most prominent class of special functions that arise when computing multi-loop integrals, and many results for Feynman integrals can be expressed in terms of them. Despite their success, it has been known since the early days of Quantum Field Theory that not all the integrals that arise in perturbative Quantum Electrodynamics can be expressed in terms of MPLs, but generalisations related to elliptic curves are needed~\cite{Sabry}. In recent years there has been a lot of activity in trying to understand Feynman integrals that evaluate to elliptic functions~\cite{Broadhurst,Bauberger:1994by,Bauberger:1994hx,Laporta:2004rb,Kniehl:2005bc,Brown:2010bw,Remiddi:2013joa,Adams:2013kgc,Adams:2016xah,Bogner:2018uus,Bloch:2013tra,Bloch:2016izu,Remiddi:2017har,Primo:2016ebd,Passarino:2016zcd,Adams:2017ejb,Chen:2017soz,Broedel:2018iwv,Ablinger:2017bjx,Broedel:2017kkb,Broedel:2017siw,Broedel:2018rwm}.
In particular, it was realised that functions of elliptic type also show up in two-loop computations relevant to LHC processes, like for example
top-quark pair production~\cite{Czakon:2008ii,vonManteuffel:2017hms,Hidding:2017jkk,Adams:2018bsn}, Higgs production~\cite{Bonciani:2016qxi,Mistlberger:2018etf}
and top-mass effects in diphoton and dijet production~\cite{Becchetti:2017abb}.  Recently it was shown that
elliptic integrals also show up in planar $\mathcal{N}=4$ Super Yang-Mills~\cite{CaronHuot:2012ab,Nandan:2013ip,Bourjaily:2017bsb,Bourjaily:2018ycu}.
These results make it clear that elliptic functions are an integral part of the arsenal of special functions that appear in perturbative Quantum Field Theory. A deeper understanding of these special functions is thus highly desirable. 

\section{A lightning review of elliptic polylogarithms}
From the mathematical point of view one expects that (part of) the class of special functions relevant for Feynman integrals are generalisations of MPLs to elliptic curves, known as \emph{elliptic multiple polylogarithms} (eMPLs)~\cite{BrownLevin}. Here, we use the following definition of eMPLs as iterated integrals on (the universal cover of) a torus, which is closely related to the eMPLs that appear in mathematics~\cite{BrownLevin} and string theory~\cite{Broedel:2014vla},
\begin{equation}\label{eq:gamt_def}
\gamtt{n_1 &\ldots& n_k}{z_1 & \ldots & z_k}{z}{\tau} = \int_0^zdz'\,g^{(n_1)}(z'-z_1,\tau)\,\gamtt{n_2 &\ldots& n_k}{z_2 & \ldots & z_k}{z'}{\tau}\,,
\end{equation}
where the $z_i$ are complex numbers and $n_i\in \mathbb{N}$ are positive integers. The integration kernels $g^{(n)}(z,\tau)$ have at most simple poles and are defined through a generating series known as the \emph{Eisenstein-Kronecker series},
\begin{equation}\label{eq:Eisenstein-Kronecker}
\frac{1}{\alpha}\,\sum_{n\ge0}g^{(n)}(z,\tau)\,\alpha^n = \frac{\theta'_1(0,\tau)\,\theta_1(z+\alpha,\tau)}{\theta_1(z,\tau)\,\theta_1(\alpha,\tau)}\,,
\end{equation}
where $\theta_1$ is the odd Jacobi theta function, and $\theta'_1$ is its derivative with respect to its first argument. 

Let us comment on the variable $\tau$ that appears in eq.~\eqref{eq:gamt_def}. It is a complex number with $\textrm{Im }\tau>0$ (i.e., it lives in the upper half-plane $\mathbb{H}=\{\tau\in\mathbb{C}:\textrm{Im }\tau>0\}$) that parametrises the `shape' of the torus. Note that different values of $\tau$ may correspond to the same torus. The redundancy is parametrised by the \emph{modular group} $SL(2,\mathbb{Z})$, which acts on $\tau$ via M\"obius transformations. In other words, if $\left(\begin{smallmatrix}a & b\\ c&d\end{smallmatrix}\right)\in SL(2,\mathbb{Z})$, then $\tau$ and $\tau'=\frac{a\tau+b}{c\tau+d}$ describe the same torus. We refer to such a M\"obius transformtion as a \emph{modular transformation}. From this we see that the space of all geometrically-distinct tori, called the \emph{moduli space of tori}, is obtained as the quotient $\mathbb{H}/SL(2,\mathbb{Z})$ of the upper half-plane $\mathbb{H}$ by the modular group $SL(2,\mathbb{Z})$.

While it is known that every elliptic curve is isomorphic to a torus, in practical applications it does usually not present itself in this form. Rather, in applications one encounters elliptic curves defined as the set of points $(x,y)$ that satisfy a polynomial equation of the form $y^2=P(x)$, where $P(x)$ is a polynomial of degree 3 or 4. We have thus two different ways of describing an elliptic curve: either as a torus with coordinate $z$ or as the set of points with coordinates $(x,y)$ subject to the polynomial constraint $y^2=P(x)$. In the case of a quartic polynomial, $P(x)=(x-a_1)\ldots(x-a_4)$, the map from $(x,y)$ to $z$ is given by \emph{Abel's map},
\begin{equation}\label{eq:Abel}
z(x,\vec a)  = \frac{c_4}{\omega_1}\int_{a_1}^x\frac{dx'}{\sqrt{P(x')}}\,.
\end{equation}
Here $c_4=\frac{1}{2}\sqrt{(a_1-a_3)(a_2-a_4)}$ and the vector $\vec a=(a_1,\ldots,a_4)$ collects the roots of the polynomial $P(x)$ and defines the `shape' of the elliptic curve. It is related to the parameter $\tau$ in eq.~\eqref{eq:gamt_def} by $\tau=\omega_2/\omega_1$, where $\omega_i$ are the \emph{periods} of the elliptic curve,
\begin{equation}
\omega_1 = 2\,\textrm{K}(\lambda) \textrm{~~and~~}\omega_2 = 2i\,\textrm{K}(1-\lambda) \,, \qquad \lambda = \frac{(a_1-a_4)(a_2-a_3)}{(a_1-a_3)(a_2-a_4)}\,.
\end{equation}
One can show that any pair of linearly independent periods uniquely defines an elliptic curve.
In this context the aforementioned $SL(2,\mathbb{Z})$ redundancy simply corresponds to a rotation of the basis of periods, i.e., the periods $(\omega_2,\omega_1)^T$ and  $(\omega'_2,\omega'_1)^T$ describe the same elliptic curve if there is $\left(\begin{smallmatrix}a & b\\ c&d\end{smallmatrix}\right)\in SL(2,\mathbb{Z})$ such that
\begin{equation}
\left(\begin{array}{c}\omega'_2\\\omega'_1\end{array}\right) = \left(\begin{array}{cc}a & b\\ c&d\end{array}\right)\left(\begin{array}{c}\omega_2\\\omega_1\end{array}\right)\,.
\end{equation}

We can also describe the eMPLs defined in eq.~\eqref{eq:gamt_def} in terms of the coordinates $(x,y)$ that define the elliptic curve. This representation is often more directly related to the Feynman integral one wants to compute. Following ref.~\cite{Broedel:2017kkb}, we define,
\begin{equation}\label{eq:E4_def}
\textrm{E}_4\left(\begin{smallmatrix}n_1 &\ldots& n_k \\c_1 & \ldots & c_k\end{smallmatrix};x,\vec a\right) = \int_0^xdx'\,\psi_{n_1}(c_1,x')\,\textrm{E}_4\left(\begin{smallmatrix}n_2 &\ldots& n_k \\c_2 & \ldots & c_k\end{smallmatrix};x',\vec a\right)\,,
\end{equation}
with $n_i\in\mathbb{Z}$, $c_i\in\mathbb{C}\cup\{\infty\}$. 
The integration kernels $\psi_{n}(c,x)$ have at most simple poles. Their explicit form quickly becomes rather involved, and we content ourselves here to present a few explicit cases that will be needed in subsequent sections. We have for example,
\begin{equation}\label{eq:psis}
\psi_0(0,x) = \frac{c_4}{y}\,,\qquad \psi_1(c,x) = \frac{1}{x-c}\,,\qquad \psi_{-1}(c,x) = \frac{y_c}{y(x-c)}\,,
\end{equation}
with $y=\sqrt{P(x)}$, $y_c=\sqrt{P(c)}$. We note that $\psi_1(c,x)$ corresponds to the kernel that defines ordinary MPLs in eq.~\eqref{eq:MPLs_def}. Hence ordinary MPLs are a subset of eMPLs. Although not at all obvious, it can be shown that the kernels $\psi_{n}(c,x)$ are in one-to-one correspondence with the kernels $g^{(n)}(z-z_c)$ that appear in eq.~\eqref{eq:gamt_def}~\cite{Broedel:2017kkb}. In particular, the kernels in eq.~\eqref{eq:psis} can be written as
\begin{equation}\begin{split}\label{eq:psi_to_g}
dx\,\psi_0(0,x) &\,= \omega_1\,dz\,,\\
dx\,\psi_1(c,x) &\,=dz\,\left[g^{(1)}(z-z_c,\tau) + g^{(1)}(z+z_c,\tau) - g^{(1)}(z-z_\infty,\tau) - g^{(1)}(z+z_\infty,\tau)\right]\,,\\
dx\,\psi_{-1}(c,x) &\,=dz\,\left[g^{(1)}(z-z_c,\tau) - g^{(1)}(z+z_c,\tau) + g^{(1)}(z_c-z_\infty,\tau) + g^{(1)}(z_c+z_\infty,\tau)\right]\,,
\end{split}\end{equation}
where $z_x \equiv z(x,\vec a)$ is the image of $x$ under the map in eq.~\eqref{eq:Abel}. A more detailed discussion of eMPLs and their properties is beyond the scope of these proceedings. We only mention here that eMPLs share many of the properties of ordinary MPLs. In particular, they form a shuffle algebra, and the algebra of eMPLs is closed under integration~\cite{BrownLevin,Broedel:2017kkb}. Moreover, using results from ref.~\cite{BrownNotes} one can define a coaction and a notion of symbols on eMPLs~\cite{Broedel:2018iwv}.

The previous discussion makes it clear that eMPLs may present themselves in different guises: they can equally-well be defined as iterated integrals $\widetilde{\Gamma}$ on a torus, or as iterated integrals $\textrm{E}_4$ on the elliptic curve defined by the polynomial equation $y^2=P(x)$. There is another representation that prominently features in both pure mathematics and applications in physics. This representation can be motivated as follows: eMPLs are functions of the `shape' of the elliptic curve, parametrised by the variable $\tau$ in eq.~\eqref{eq:gamt_def}. In physics applications $\tau$ will depend on the external kinematic data, and it is therefore natural to expect that a dual representation of eMPLs exists as iterated integrals in the parameter $\tau$, i.e., as iterated integrals on the {moduli space}. In general, the corresponding class of integrals is much less studied in the literature, and we therefore restrict ourselves to a special case where all the arguments $z_i$ and $z$ in eq.~\eqref{eq:gamt_def} are rational points of the form $\frac{r}{N}+\tau\frac{s}{N}$, with $r,s,N$ integers. While it may seem that this special case is very restrictive, it is known that the sunrise and the kite integrals can be reduced to this case. It can be shown that eMPLs evaluated at rational points can be written as linear combinations of {iterated integrals} of the form~\cite{Broedel:2018iwv},
\begin{equation}\begin{split}
I\left(\begin{smallmatrix} n_1& N_1\\ r_1& s_1\end{smallmatrix}\big|\ldots\big|\begin{smallmatrix} n_k& N_k\\ r_k& s_k\end{smallmatrix};\tau\right) &\,\equiv  \int_{i\infty}^{\tau}d\tau'\,h_{N_1,r_1,s_1}^{(n_1)}(\tau')\,I\left(\begin{smallmatrix} n_2& N_2\\ r_2& s_2\end{smallmatrix}\big|\ldots\big|\begin{smallmatrix} n_k& N_k\\ r_k& s_k\end{smallmatrix};\tau'\right)\,, \label{eq:itmod}
\end{split}\end{equation}
where the integration kernels are Eisenstein series,
\begin{equation}\label{eq:Eisenstein}
h^{(n)}_{N,r,s}(\tau) = \sum_{k=0}^n\frac{(2\pi i s)^k}{k!\,N^k} g^{(n-k)}\left(\frac{r}{N}+\tau\frac{s}{N};\tau\right)=\!\!\!\!\!\!\!\sum_{\substack{(\alpha,\beta)\in \mathbb{Z}^2\\ (\alpha,\beta)\neq(0,0)}}\frac{e^{-2\pi i(s\alpha-r\beta)/N}}{(\alpha+\beta\tau)^{n}}\,.
\end{equation}
The integers $N$ and $n$ are called the \emph{level} and the \emph{weight} of the Eisenstein series.
The iterated integrals in eq.~\eqref{eq:itmod} are a special case of more general iterated integrals of modular forms considered in the mathematics literature~\cite{Manin,BrownModular}, though a detailed discussion of modular forms and Eisenstein series would go beyond the scope of these proceedings. Here it suffices to define a modular form $f(\tau)$ of weight $n$ for a group $\Gamma\subseteq SL(2,\mathbb{Z})$ as a holomorphic function on the complex upper half-plane that transform nicely under M\"obius transformations for $\Gamma$,
\begin{equation}
f\left(\frac{a\tau+b}{c\tau+d}\right) = (c\tau+d)^n\,f(\tau)\,,\qquad \left(\begin{smallmatrix}a&b \\c&d\end{smallmatrix}\right)\in\Gamma\,.
\end{equation}
One can then show that the Eisenstein series defined in eq.~\eqref{eq:Eisenstein} are modular forms for the \emph{congruence subgroup} $\Gamma(N)\equiv\left\{\gamma\in SL(2,\mathbb{Z}): \gamma = \left(\begin{smallmatrix}1&0 \\0&1\end{smallmatrix}\right) \!\!\!\!\!\!\mod N\right\}$~\cite{Broedel:2018iwv}.
In other words, and a bit over-simplifying, we can think of modular forms as holomorphic functions with nice transformation properties defined on the moduli space of tori. 

Based on the previous discussion, we conclude that eMPLs evaluated at rational points can be represented equivalently in terms of three different classes of iterated integrals:
\begin{enumerate}
\item as eMPLs $\widetilde{\Gamma}$ in eq.~\eqref{eq:gamt_def} evaluated at rational points $z_i = \frac{r_i}{N_i}+\tau\frac{s_i}{N_i}$.
\item as eMPLs E$_4$ defined in eq.~\eqref{eq:E4_def}.
\item as iterated integrals of Eisenstein series defined in eq.~\eqref{eq:itmod}.
\end{enumerate}

In the remainder of these proceedings we will illustrate this threefold way of representing eMPLs on the example of a novel class of special functions introduced by Remiddi and one of the authors in ref.~\cite{Remiddi:2017har} in the context of the sunrise integral. We start by reviewing the functions of ref.~\cite{Remiddi:2017har} in the next section. The main novel result introduced in these proceedings is a proof that the functions of ref.~\cite{Remiddi:2017har} can always, and very naturally, be described in terms of eMPLs and iterated integrals of modular forms. 

\section{An elliptic generalisation of MPLs}
In ref.~\cite{Remiddi:2017har} a new class of special functions was introduced. 
The definition of this new class of function is
\begin{equation}\label{eq:EG_def}
\textrm{EG}^{[n]}(k,u) = \int_{b_i}^{b_j}\frac{db\,b^k}{\sqrt{R_4(u,b)}}\,G(b_{j_1},\ldots,b_{j_n};b)\,,
\end{equation}
where $R_4(u,b) = (b-b_1)\ldots (b-b_4)$, with
\begin{equation}
b_1 = 0\,,\quad b_2 = 4\,,\quad b_3 = (\sqrt{u}-1)^2\,,\quad b_4 = (\sqrt{u}+1)^2\,.
\end{equation}
These functions appear in the computation of the two-loop sunrise integral with three equal masses, with $u=p^2/m^2$, $p$ being the external momentum. A detailed study of some of the properties of these functions, in particular relations in low weight $n$ coming from integration-by-parts identities, have been studied in detail in ref.~\cite{Remiddi:2017har}. Here we follow a different route, and we show that the functions defined by eq.~\eqref{eq:EG_def} can naturally be expressed in terms of eMPLs, and they can therefore be written in any of the three ways discussed at the end of the previous section. For simplicity, we only discuss a single example (cf. eq. (5.16) of ref.~\cite{Remiddi:2017har}), 
\begin{equation}
\mathcal{I}(u) = \int_{b_2}^{b_3}\frac{db}{\sqrt{R_4(u,b)}} \log(b-4) = \int_{b_2}^{b_3}\frac{db}{\sqrt{R_4(u,b)}} \left[G(b_2,b)+2\log2-i\pi\right]\,.
\end{equation}
In the remainder of these proceedings we will discuss in some detail how this integral can be represented in terms of each of the three classes of functions defined in the previous section. We stress, however, that all results of the next subsections generalise to arbitrary representatives of the functions defined in eq.~\eqref{eq:EG_def}.

\subsection{Representation in terms of eMPLs E$_4$}
It is straightforward to write $\mathcal{I}(u)$ in terms of the eMPLs E$_4$. Indeed, we see from eq.~\eqref{eq:psis} that ordinary MPLs are a subset of eMPLs, and we have
\begin{equation}
G(b_2;b) = \int_0^b\frac{db'}{b'-b_2} = \int_0^bdb'\,\psi_1(b_2,b) = \textrm{E}_4\left(\begin{smallmatrix} 1\\b_2\end{smallmatrix};b,\vec b\right)\,,
\end{equation}
where we have introduced the shorthand $\vec b=(b_1,\ldots,b_4)$. The remaining integration can then immediately be performed using eq.~\eqref{eq:E4_def}, after realising that $1/\sqrt{R_4(u,b)} = 1/c_4\,\psi_0(0,b)$, with $c_4 = \frac{1}{2}\sqrt{(b_1-b_3)(b_2-b_4)}$. We find
\begin{align}\label{eq:I_E4}
\mathcal{I}(u) &\,= \frac{1}{c_4}\int_{b_2}^{b_3}db\,\psi_0(0,b)\left[\textrm{E}_4\left(\begin{smallmatrix} 1\\b_2\end{smallmatrix};b,\vec b\right)+ 2\log2 - i\pi \right]\\
&\,=\frac{1}{c_4}\,\textrm{E}_4\left(\begin{smallmatrix} 0&1\\0&b_2\end{smallmatrix};b_3,\vec b\right) -\frac{1}{c_4}\,\textrm{E}_4\left(\begin{smallmatrix} 0&1\\0&b_2\end{smallmatrix};b_2,\vec b\right) + \frac{1}{c_4}\left(2\log2-i\pi\right)\, \left[\textrm{E}_4\left(\begin{smallmatrix} 0\\0\end{smallmatrix};b_3,\vec b\right)-\textrm{E}_4\left(\begin{smallmatrix} 0\\0\end{smallmatrix};b_2,\vec b\right)\right] \,. 
\nonumber
\end{align}

\subsection{Representation in terms of eMPLs $\widetilde{\Gamma}$}
We now show how the expression in terms of E$_4$ integrals can be converted to the eMPLs $\widetilde{\Gamma}$ defined on the torus. We start by using eq.~\eqref{eq:Abel} to map each of the points relevant to our problem to the torus. We find, with $z_i\equiv z(b_i,\vec b)$ and $z_{\infty}\equiv z(\infty,\vec b)$,
\begin{equation}\label{eq:punctures}
z_1 = 0\,,\quad z_2 = \frac{\tau}{2}\,,\quad  z_3 = \frac{1}{2}+\frac{\tau}{2}\,,\quad z_4 = \frac{1}{2}\,,\quad z_{\infty} = \frac{1}{3}\,.
\end{equation}
We can now use eq.~\eqref{eq:psi_to_g} to change the integration variable from $x$ to $z \equiv z(x,\vec b)$. We illustrate this procedure only on one of the E$_4$ functions appearing in eq.~\eqref{eq:I_E4}, though all other cases work in a similar way. We find
\begin{align}
&\textrm{E}_4\left(\begin{smallmatrix} 0&1\\0&b_2\end{smallmatrix};b_3,\vec b\right) = \int_0^{b_3}db\,\psi_0(0,b)\int_0^{b}db'\,\psi_1(b_2,b')\\
\nonumber&= \omega_1\int_0^{1/2+\tau/2}dz\int_0^{z}dz'\,\left[g^{(1)}(z'-\tau/2,\tau) + g^{(1)}(z+\tau/2,\tau) - g^{(1)}(z-1/3,\tau) - g^{(1)}(z+1/3,\tau)\right]\\
\nonumber&= \omega_1\!\!\left[\gamtt{0&1}{0&\tau/2}{\frac{1}{2}+\frac{\tau}{2}}{\tau}+\gamtt{0&1}{0&-\tau/2}{\frac{1}{2}+\frac{\tau}{2}}{\tau}
-\gamtt{0&1}{0&1/3}{\frac{1}{2}+\frac{\tau}{2}}{\tau}-\gamtt{0&1}{0&-1/3}{\frac{1}{2}+\frac{\tau}{2}}{\tau}\right]\!\!.
\end{align}
If we repeat these steps for all the E$_4$ functions in eq.~\eqref{eq:I_E4}, we find
\begin{equation}\begin{split}\label{eq:I_sGt}
&\mathcal{I}(u) = \frac{\omega_1}{c_4}\left[\gamtt{0&1}{0&\tau/2}{\frac{1}{2}+\frac{\tau}{2}}{\tau}+\gamtt{0&1}{0&-\tau/2}{\frac{1}{2}+\frac{\tau}{2}}{\tau}-\gamtt{0&1}{0&\tau/2}{\frac{\tau}{2}}{\tau}-\gamtt{0&1}{0&-\tau/2}{\frac{\tau}{2}}{\tau}\right.\\
&\,\left.-\gamtt{0&1}{0&1/3}{\frac{1}{2}+\frac{\tau}{2}}{\tau} +\gamtt{0&1}{0&1/3}{\frac{\tau}{2}}{\tau}-\gamtt{0&1}{0&-1/3}{\frac{1}{2}+\frac{\tau}{2}}{\tau} +\gamtt{0&1}{0&-1/3}{\frac{\tau}{2}}{\tau} +\log2 -\frac{i\pi}{2}\right]\,.
\end{split}\end{equation}

\subsection{Representation in terms of iterated integrals of modular forms}
From eq.~\eqref{eq:punctures} we see that all the arguments of the eMPLs are rational points on the torus of the form $\frac{r}{6}+\tau\frac{s}{6}$, and so it must be possible to write $\mathcal{I}(u)$ as iterated integrals of Eisenstein series $h^{(n)}_{6,r,s}(\tau)$ for the group $\Gamma(6)$~\cite{Broedel:2018iwv}. 
The main idea that allows us to achieve this is to compute the derivative with respect to $\tau$ of each eMPL, and to integrate back to obtain an iterated integral representation in $\tau$. A closed formula for the total derivative of an eMPL was presented in ref.~\cite{Broedel:2018iwv}. The derivative will lower the length, i.e., the number of iterated integrations, by one unit, and we can proceed recursively. We will  only discuss an example of an eMPL of length one in detail, and we focus on $\gamtt{1}{1/3}{\frac{\tau}{2}}{\tau}$. Using the results of ref.~\cite{Broedel:2018iwv}, we can easily compute the derivative,
\begin{equation}\begin{split}
\partial_{\tau}\gamtt{1}{1/3}{\frac{\tau}{2}}{\tau} &\,= \frac{1}{2}\,g^{(1)}\left(\frac{1}{6}+\frac{\tau}{2},\tau\right) - \frac{1}{2\pi i}\,g^{(2)}\left(\frac{1}{3},\tau\right) + \frac{1}{2\pi i}\,g^{(2)}\left(\frac{1}{6}+\frac{\tau}{2},\tau\right)\\
&\,=\frac{1}{2\pi i}\,h^{(2)}_{6,1,3}(\tau) - \frac{1}{2\pi i}\,h^{(2)}_{6,2,0}(\tau) - \frac{i\pi}{4}\,,
\end{split}\end{equation}
where in the last step we used eq.~\eqref{eq:Eisenstein} to rewrite the integration kernels $g^{(n)}$ evaluated at rational points in terms of Eisenstein series. We can then immediately write
\begin{equation}
\gamtt{1}{1/3}{\frac{\tau}{2}}{\tau} = C + \frac{1}{2\pi i}\,I\left(\begin{smallmatrix} 2&6\\1&3\end{smallmatrix};\tau\right) - \frac{1}{2\pi i}\,I\left(\begin{smallmatrix} 2&6\\2&0\end{smallmatrix};\tau\right) - \frac{i\pi}{4}\,I\left(\begin{smallmatrix} 0&0\\0&0\end{smallmatrix};\tau\right)\,,
\end{equation}
where $C$ is an integration constant that can be obtained by analysing the behaviour of $\gamtt{1}{1/3}{\frac{\tau}{2}}{\tau}$ as $\tau\to i\infty$. This can easily be done since the integration kernels $g^{(n)}$ become simpler in the limit, and all integrations can be easily performed~\cite{Broedel:2014vla,Broedel:2015hia,Broedel:2017jdo}. We find
\begin{equation}
\lim_{\tau\to i\infty}\gamtt{1}{1/3}{\frac{\tau}{2}}{\tau} = -\frac{i\pi}{2}\tau-\frac{2\pi i}{3}-\frac{1}{2}\log3\,.
\end{equation}
Putting everything together, we arrive at the final expression for $\mathcal{I}(u)$ in terms of iterated integrals of Eisenstein series for $\Gamma(6)$,
\begin{equation}
\mathcal{I}(u) = \frac{1}{2\pi i}\,I\left(\begin{smallmatrix} 2&6\\2&0\end{smallmatrix};\tau\right)-\frac{1}{2\pi i}I\left(\begin{smallmatrix} 2&6\\0&3\end{smallmatrix};\tau\right)+\frac{i\pi}{4}\tau+\log2+\frac{1}{2}\log3\,.
\end{equation}

\section{Conclusions}
In these proceedings we have reviewed several classes of special functions that feature prominently in pure mathematics and in perturbative Quantum Field Theory, namely elliptic multiple polylogarithms and iterated integrals of Eisenstein series. We have shown that these classes of functions are intimately related in the case where all the arguments of the eMPLs are rational points, and we have discussed how one can switch from one representation to the other. 

As a novel application of our formalism, we have shown that the elliptic generalisations of MPLs introduced in ref.~\cite{Remiddi:2017har} can naturally be expressed in terms of eMPLs. When mapped to the torus, we see that all arguments of the eMPLs are given by the set of rational points in eq.~\eqref{eq:punctures}. Consequently, the functions introduced in ref.~\cite{Remiddi:2017har} can equally-well be expressed in terms of iterated integrals of Eisenstein series for $\Gamma(6)$. 

We foresee that eMPLs and iterated integrals of modular forms will play an important role in the context of precision computations in high-energy physics in the future. Gaining a deeper understanding of the mathematical properties of these functions, in particular the role played in the context of differential equations satisfied by Feynman integrals and how to evaluate them numerically in a fast and efficient way, will therefore be one of the major challenges for mathematical physics in the years to come. We believe that this line of research will benefit from a fruitful exchange between mathematics and physics that in the long run will have a positive impact on both fields.

\section*{Acknowledgments} 
This research was supported by the the ERC grant 637019 ``MathAm'', and the U.S.
Department of Energy (DOE) under contract DE-AC02-76SF00515.

\end{document}